\documentclass[11pt,a4paper]{article}

\usepackage[a4paper,margin=1in]{geometry}
\usepackage{microtype}
\usepackage{graphicx}
\usepackage{subcaption}
\usepackage{booktabs}
\usepackage{longtable}
\usepackage{amsmath}
\usepackage{amssymb}
\usepackage{enumitem}
\usepackage[numbers,sort&compress]{natbib}
\usepackage{url}
\usepackage{caption}
\usepackage{placeins}
\usepackage{xcolor}
\definecolor{linkblue}{HTML}{1A4F9C}
\usepackage[colorlinks=true,allcolors=linkblue]{hyperref}

\newcommand{\hof}{HoF-Bench}
\newcommand{\nano}{\texttt{nano-analyzer}}
\newcommand{\code}[1]{\texttt{#1}}

\title{HoF-Bench: Rediscovering Real AI-Discovered CVEs Without Frontier Models}

\author{
\small
Petr Simecek\thanks{Corresponding author:
\href{mailto:petr.simecek@aisle.com}{\texttt{petr.simecek@aisle.com}}.
Also affiliated with the Central European Institute of Technology,
Masaryk University, Brno, Czech Republic.},
Elnaz Babayeva,
Jiri Balhar,
Michal Bida,
Michal Buran,\\
\small
Vaclav Cadek,
Luigino Camastra,
Tomas Dulka,
Michal Janocko,
Tomas Klohna,\\
\small
Pavel Kohout,
Ondrej Kokes,
Adam Krivka,
Jakub Kubik,
Patrik Mada,
Igor Morgenstern,\\
\small
Marek Pavelka,
Joshua Rogers,
Petr Stastny,
Jan Tattermusch,
Dmitrijs Trizna,\\
\small
Martin Votruba,
Guido Vranken,
Jakub Zikl,
Evelina Gabasova, and
Stanislav Fort\\[0.6em]
\small AISLE, San Francisco, CA, USA and Prague, Czech Republic}
\date{July 29, 2026}

\begin{document}
\maketitle

\begin{abstract}
LLM-based analyzers have begun finding real vulnerabilities in mature open-source projects:
AISLE's analyzer is credited with more than 280 CVEs across 78 projects, including OpenSSL, curl, and GnuTLS.
We introduce \hof\ (named after AISLE's public Hall of Fame), a benchmark built from 95 of these public AI-discovered CVEs across eight repositories pinned at vulnerable commits.
Analyzers receive source and target-file scope but not CVE identifiers, descriptions, fixes, or expected mechanisms; a detector-blinded frontier-model judge credits only findings that identify the same code path, root cause, attack condition, and impact.
A deliberately minimal LLM-based analyzer rediscovers up to 65 of the 95 CVEs (68\%) under this strict protocol.
No frontier model performs detection anywhere in the study.
The ten detector backbones are five open-weight models (21B--284B total parameters, 3--13B active) and five proprietary small or ``flash''-tier models.
All of them run in the fixed scaffold with four repeated passes, an optional generated-context stage, and a replayable multi-round triage stage (7,600 model--CVE pass records).
Difficulty is strongly structured by language; the CVEs missed by every model concentrate in C infrastructure code.
\hof\ provides a compact test bed for comparing vulnerability scanners, their reliability across repeated runs, and the candidate volume they create.
The dataset is available at \url{https://huggingface.co/datasets/aisleinc/HoF-Bench}.
\end{abstract}

\section{Introduction}

Systems based on large language models (LLMs) have moved from explaining and exploiting known vulnerabilities toward finding and reporting new ones.
AISLE's LLM-based analyzer has been credited with CVEs in heavily reviewed projects, including OpenSSL, curl, GnuTLS, and widely deployed web applications \citep{aisle2026cves}.
These are repositories that had already received years of human review, fuzzing, and commercial static analysis, yet the bugs were still there.
That progress raises an obvious evaluation question: if an LLM analyzer found a vulnerability once, can another analyzer find it again from the vulnerable source?
Many security benchmarks evaluate snippet classification, patch generation, exploit reproduction, or purpose-built vulnerable applications.
Those tasks are valuable, but they do not directly measure whether a scanner report identifies the CVE-worthy bug in a real, complex software setting.

A \emph{CVE} is a public identifier for a disclosed vulnerability.
A scanner emits \emph{findings}: structured claims about vulnerable code, attacker control, and impact.
\emph{Triage} is the follow-up review that rejects weak findings before human review.
Traditional static application security testing (SAST) usually applies program-analysis rules that look for common patterns, without the generative reasoning used by LLMs.

\hof\ takes its name from AISLE's public Hall of Fame and asks that question on 95 recent public CVEs attributed by AISLE to its systems and listed there \citep{aisle2026cves}.
We call the task \emph{source-conditioned} because the benchmark supplies the vulnerable target files: it tests analysis and rediscovery, not autonomous repository-wide localization.
As of July 2026, AISLE's Hall of Fame listed 287 assigned CVEs across 78 projects, including 134 with High or Critical severity.
We concentrate on eight source-code repositories with at least five discoveries each that could be grounded at a common vulnerable snapshot, yielding 95 CVEs.
With one pinned commit per repository, a tool can run the benchmark using only eight checkouts and a scanner-visible manifest.
Scoring is more demanding: credit requires strict vulnerability identity; file proximity or vulnerability class alone is insufficient.

Our quantitative question is how much of this population a non-frontier LLM analyzer can rediscover and which system choices affect the result.
To answer that, we use a reference scanner inspired by AISLE's already open-sourced, intentionally minimal \nano\ project \citep{aisle2026nano}.
The detector panel comprises five open-weight mixture-of-experts models with 3--13B active parameters and five proprietary ``flash''-tier models, priced between \$0.03 and \$1.50 per million input tokens at run time.
Holding the scanner scaffold fixed while varying detector models and evaluation conditions lets us ask practical questions that a single benchmark score hides: which detector models rediscover the largest fraction of CVEs, how much repeated scanning helps, whether generated context and deeper triage add signal or mostly add noise and cost, how many deduplicated finding clusters are produced per rediscovered CVE, and which CVEs are consistently easy or hard across models.

The short answer is that most real AI-found CVEs can be rediscovered, but not by any single run.
Six of the ten evaluated model families rediscover at least 50 of the 95 CVEs in one of the four-pass conditions, and the highest observed count is 65.
Among the pipeline choices we evaluate, repeated scanning produces the most consistent gain in cumulative rediscovery.
Averaged over all pass subsets, four passes find approximately 12--20 more CVEs per model than one pass.
Distributing the same number of passes across different models improves rediscovery further because their blind spots only partially overlap.
The two most natural-looking pipeline additions matter much less:
Generated context reduces the number of CVEs missed by every model, but its pooled recall effect is small and uncertain while candidate volume rises.
Increasing triage from one round to three rounds plus an arbiter shifts strict outcomes by only a few percent.
Finally, difficulty varies widely across CVEs: with context, 21 are found by every model and 11 by none (22 and 16 without context).
Both extremes are structured by language: 18 of the 21 universally found CVEs are JavaScript or PHP, while nine of the 11 universal misses are in C.

Our main contributions in this paper are three.
The benchmark itself, \hof\ v1, holds 95 public CVEs across eight pinned repositories with scanner-visible and evaluator-only fields separated.
A strict rediscovery protocol is built around a detector-blinded automated judge.
A controlled study then evaluates ten detector models under repeated passes and context on/off.
The study totals 7,600 analyzer runs (95 CVEs $\times$ 10 models $\times$ 2 context conditions $\times$ 4 passes); triage-depth variants are evaluated by replaying stored reviews, with full failure accounting.

\section{HoF-Bench v1}

\subsection{Construction}

\begin{figure}[!htbp]
  \centering
  \includegraphics[width=0.82\linewidth]{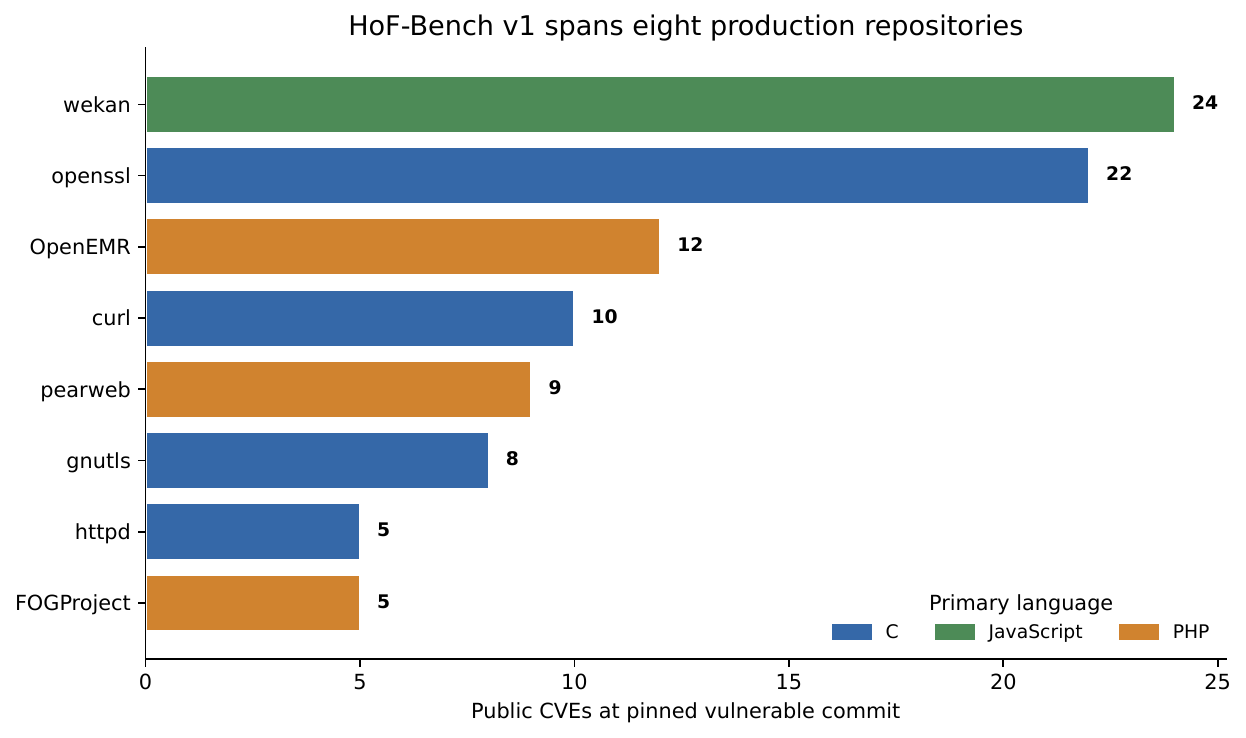}
  \caption{\hof\ contains 95 public CVEs across eight repositories and three primary languages.}
  \label{fig:composition}
\end{figure}

\hof\ v1 starts from public CVEs attributed by AISLE to its systems \citep{aisle2026cves}.
From this population, we selected repositories with at least five AISLE-attributed disclosures and a manageable common vulnerable snapshot.
We kept all validated cases in four C repositories, and the application cases that could be confidently grounded at one commit in four PHP/JavaScript repositories; excluded records could not be tied to the repository, could not be verified at the common commit, or would have required a different snapshot.
The result is 95 CVEs across eight repositories (OpenSSL, curl, GnuTLS, Apache httpd, pearweb, FOGProject, OpenEMR, and WeKan), of which 31 are multi-file cases.
Each repository has one scan commit, so external tools can be evaluated with only eight checkouts.
The benchmark spans 39 CWE identifiers.
The most frequent are cross-site scripting (CWE-79; 9 cases), SQL injection (CWE-89; 8), incorrect authorization (CWE-863; 8), improper access control (CWE-284; 7), and NULL-pointer dereference (CWE-476; 6).
More broadly, the cases include authorization and authentication failures, injection and cross-site scripting, denial-of-service and memory-safety bugs, information disclosure, and certificate or cryptographic validation flaws.
These mechanisms are not distributed evenly across languages: denial-of-service and memory-safety cases concentrate in C, SQL injection in PHP, and authorization and access-control failures in the web applications.
Figure~\ref{fig:composition} shows the benchmark composition.

\subsection{Task Definition}

Each task gives the scanner access to the full repository at a pinned commit and designates 1--7 target files for vulnerability analysis.
This target-file scope is derived from the benchmark ground truth and includes the primary vulnerable file, providing a substantial localization oracle.
The designated files contain a median of 911 source lines and are included in the detector prompt, subject to a 300,000-character cap.
The context and triage stages may additionally issue narrow literal searches across the full checkout.
The scanner does not receive the CVE identifier, vulnerability description, fix commit, or expected vulnerable snippet.
The evaluator has those fields and uses them only after the scanner emits findings.
The primary outcome is strict rediscovery: the finding must identify the same vulnerable code path or component, root cause, attacker-controlled condition or trust boundary, and security impact.

\begin{figure}[!t]
  \centering
  \includegraphics[width=0.98\linewidth]{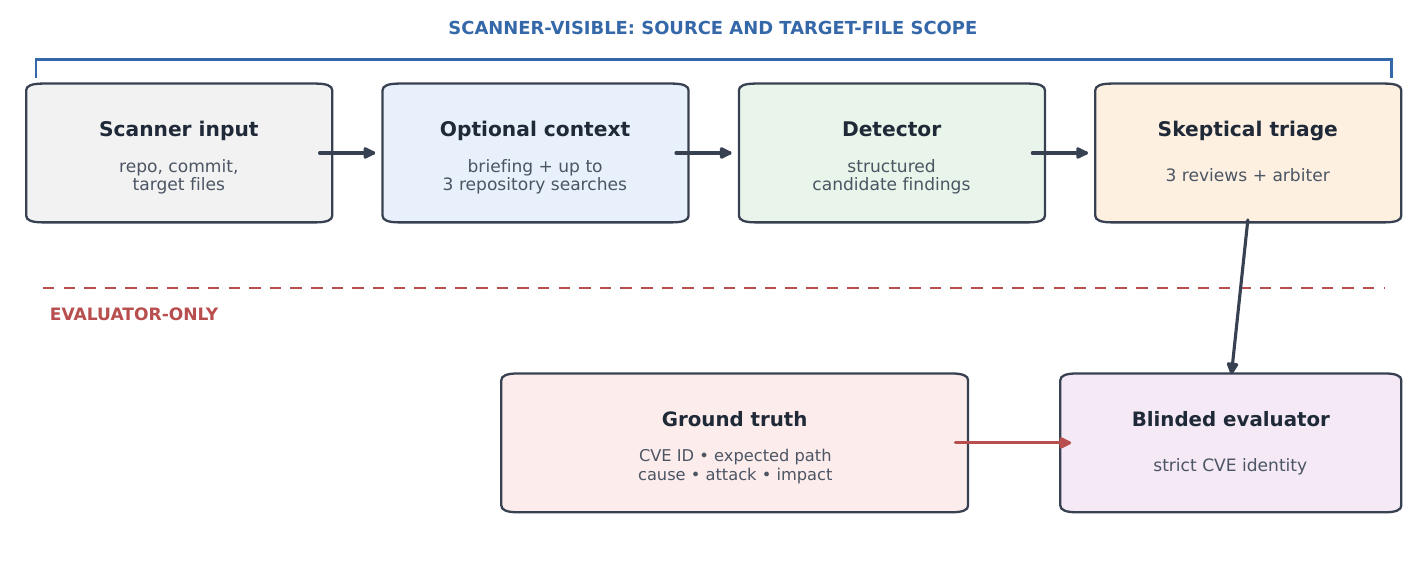}
  \caption{Reference-scanner pipeline and information boundary. The detector never receives evaluator-only CVE ground truth.}
  \label{fig:pipeline}
\end{figure}

\section{Related Work}

Vulnerability-detection benchmarks typically work with code snippets, functions, or synthetic test cases rather than full production repositories.
Devign \citep{devign2019}, Big-Vul \citep{bigvul2020}, and PrimeVul \citep{primevul2025} focus on functions or code changes; Juliet \citep{juliet} supplies synthetic tests; SARD \citep{sard} spans synthetic programs to production applications; and Snyk VulnBench JS 1.0 measures the repeatability of LLM review on ten small JavaScript fixtures against one SAST tool's own reference findings \citep{snykvulnbench2026}.
Such targets keep evaluation cheap and inspectable, but they rarely require the evidence that reviewers actually chase in production code: state that crosses files, configuration that decides whether a check applies, and trust boundaries that are only visible from the calling side.
\hof\ instead asks whether a scanner can rediscover a specific CVE mechanism in a pinned snapshot of the real project.

Recent benchmarks move closer to repository-level evaluation.
RepoPairBench provides 100 Python vulnerability/fix pairs in complete repositories and checks whether model rationales match the documented mechanisms \citep{sun2026drea}; JITVul links functions to introducing and fixing commits for 879 CVEs \citep{jitvul2025}.
RealVuln compares rule-based SAST, general-purpose LLMs, and security-specialized scanners on 26 intentionally vulnerable Python repositories \citep{realvuln2026}.
BountyBench and SEC-bench go further along the lifecycle, into exploitation, proof-of-concept generation, and patching \citep{bountybench2025,secbench2025}.

A complementary line of work catalogs rather than benchmarks.
The Berkeley Vulnerability Initiative's Agentic Vulnerability Coverage Map maintains a daily, multi-vendor census of CVEs attributed to agentic systems, tracking which vulnerability classes those systems are gaining and losing ground on \citep{bvi2026agentic}.
A census of that kind and a benchmark answer different questions: the census measures what agentic systems are finding in the wild, while \hof\ measures whether a given scanner can find a fixed set of them again.
It also supplies an external reference point for the population \hof\ samples, which we return to when discussing the benchmark's boundaries.

\hof\ occupies the space these leave open, between snippet classification and full exploitation: production snapshots pinned before the fix, credit only for strict CVE identity, and repeated passes so that reliability is measurable rather than assumed.
It also differs in scope and in cost of entry.
The repository-level benchmarks above are single-language, whereas \hof\ spans C, PHP, and JavaScript, and that turns out to matter: difficulty tracks language more strongly than vulnerability class (Table~\ref{tab:language}).
Running it requires eight checkouts and a manifest, which keeps the barrier low enough that a tool author can reproduce our numbers in an afternoon.
Our analysis of variability across repeated passes builds on two recent AISLE studies \citep{simecek2026silent,buran2026benchmarks}; Supplement~\ref{sec:reliability-model} describes the statistical model.

\section{Reference Scanner and Evaluation Protocol}

\subsection{Pipeline}

The reference scanner follows the broad structure described in AISLE's \nano\ blog post \citep{aisle2026nano}: optional context construction, vulnerability detection, and skeptical triage.
We use the scaffold as a controlled comparison instrument and do not claim it as a contribution.
Its purpose is to provide a transparent, fixed instrument for comparing models and repeated runs.
The benchmark can therefore be used in two ways: swapping a detector model into this scaffold, or running an external tool (conventional SAST or agentic) against the eight pinned snapshots and scoring its reports with the same judge.
Figure~\ref{fig:pipeline} summarizes the scanner stages and blinded evaluator boundary.

In context-enabled runs, a model first prepares a concise security briefing that summarizes the component's purpose, externally reachable entry points, trust boundaries, and security-relevant data and control flow, without yet claiming a vulnerability.
It may also request up to three narrow repository searches for callers, constants, or defenses; the results are appended to the briefing.
The detector receives the target source files together with this briefing.
In no-context runs, it receives the same source files with the briefing disabled.
All runs use three sequential triage rounds followed by an arbiter.
The arbiter sees the finding, source code, and all three reviews, then makes a final valid-or-invalid decision from the code evidence rather than tallying votes.
Runs also impose a maximum of five raw findings per pass and bounded structured-output retries.
Because every per-round triage verdict is stored, shallower triage policies can be replayed from the recorded reviews, which we use in the results to isolate the effect of triage depth.

\begin{table}[!h]
  \centering
  \caption{Detector panel. Parameter counts are total/active per token for the sparse mixture-of-experts open-weight models; proprietary vendors do not disclose sizes. Prices are API snapshots at run time (July 2026), USD per million tokens.}
  \label{tab:model-panel}
  \begin{tabular}{llllrr}
    \toprule
    Model & Weights & License & Params (act.) & In \$/M & Out \$/M \\
    \midrule
    GPT-OSS 20B & open & Apache 2.0 & 21B (3.6B) & 0.03 & 0.13 \\
    GPT-OSS 120B & open & Apache 2.0 & 117B (5.1B) & 0.04 & 0.17 \\
    DeepSeek V4 Flash & open & MIT & 284B (13B) & 0.09 & 0.19 \\
    Qwen 3.6 35B A3B & open & Apache 2.0 & 35B (3B) & 0.14 & 1.00 \\
    Mistral Small 4 & open & Apache 2.0 & 119B (6.5B) & 0.15 & 0.60 \\
    GPT-5.4 nano & proprietary & --- & undisclosed & 0.20 & 1.25 \\
    Gemini 2.5 Flash & proprietary & --- & undisclosed & 0.30 & 2.50 \\
    GPT-5.4 mini & proprietary & --- & undisclosed & 0.75 & 4.50 \\
    GPT-5.6 luna & proprietary & --- & undisclosed & 1.00 & 6.00 \\
    Gemini 3.5 Flash & proprietary & --- & undisclosed & 1.50 & 9.00 \\
    \bottomrule
  \end{tabular}
\end{table}

\subsection{Inexpensive, Non-Frontier Detector Models}

We do not use frontier models as detectors.
Instead, we ask how far the fast, inexpensive tier gets when placed inside a systematic scaffold.
That question comes out of an earlier, much smaller probe in which open-weight models with 3--5B active parameters recovered much of the published analysis of several frontier-model showcase vulnerabilities once the relevant code had been isolated, and in which the ranking of models reshuffled from one task to the next \citep{fort2026mythos}.
Two of the models used there, GPT-OSS 20B and GPT-OSS 120B, are in the panel below.
\hof\ asks the same question at benchmark scale and under a blinded judge.
The ten detectors, summarized in Table~\ref{tab:model-panel}, are split evenly between open-weight and proprietary families, precisely so that the benchmark demonstrates both routes.
The five open-weight models ship under MIT or Apache~2.0 licenses and range from 21B to 284B total parameters, all with sparse mixture-of-experts architectures that activate only 3--13B parameters per token; the five proprietary models are the vendors' small or ``flash'' tiers with undisclosed sizes.
API prices at run time spanned \$0.03--\$1.50 per million input tokens and \$0.13--\$9.00 per million output tokens.
For comparison, contemporary frontier tiers such as GPT-5.6 terra cost \$2.50 and \$15.00.
The only larger model anywhere in the study is GPT-5.6 terra, used exclusively as the blinded judge, never as a detector.
The observed cost--rediscovery tradeoff, normalized per benchmark CVE, is shown in Supplement Figure~\ref{fig:supp-recall-cost}.

Exact provider model identifiers are listed in the supplement and run artifacts.
All detector calls use medium reasoning effort where supported.
Models without a provider prefix were called through the OpenAI API; provider-prefixed models were called through OpenRouter.
Raw records preserve requested and returned model identifiers, request IDs, token usage, and latency.

\begin{figure}[!t]
  \centering
  \includegraphics[width=0.95\linewidth]{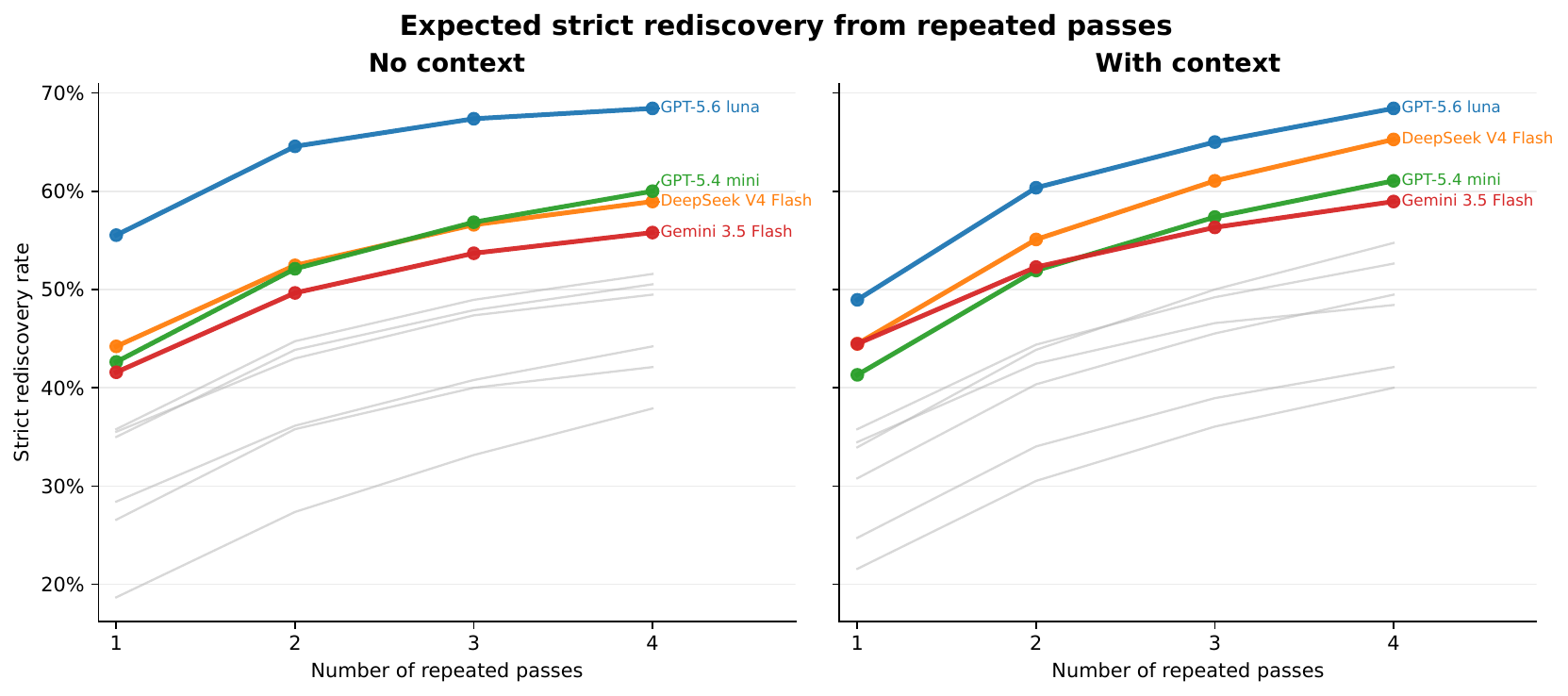}
  \caption{Expected strict rediscovery rate: the percentage of the 95 benchmark CVEs found by at least one of a random subset of the four repeated passes. The four models with the highest pass@4 recall across either condition are colored and labeled; the other six are shown in gray. Each point averages over all pass subsets of size $k$, so the curve does not depend on an arbitrary pass order.}
  \label{fig:curves}
\end{figure}

\subsection{Evaluation}

Each model is run for four repeated passes over all 95 CVEs, with and without context, for 7,600 pass records.
Surviving findings are scored by a fixed automated LLM judge that sees the task ground truth but is blinded to detector identity, context condition, and pass number.
A task is counted as rediscovered when at least one surviving finding receives a strict-match verdict.
The current implementation does not perform cross-task semantic assignment, which we list as a limitation.
The judge assigns one of four outcomes: \code{match}, \code{partial}, \code{no\_match}, or \code{insufficient\_evidence}.
We report strict recall as the primary metric and retain partial matches for sensitivity analysis.
Success@$k$ is the expected number of benchmark CVEs found at least once in a random subset of $k$ passes; dividing by 95 gives the strict rediscovery rate.
We compute it over all subsets of the four passes, avoiding dependence on an arbitrary pass ordering, and we write pass@4 for the union of all four passes.
Because every task is positive and unmatched findings are not independently labeled, this is a recall metric only.
We therefore do not call unmatched findings false positives.
To estimate human review burden, we semantically deduplicate post-triage findings by grouping reports that describe the same reviewer-actionable issue.
We do this separately for each detector model and context condition across all four passes.
Clear duplicates are identified using deterministic similarity rules, while ambiguous pairs are assessed by an LLM judge.
We then divide the number of deduplicated finding clusters by the number of benchmark CVEs that the model strictly rediscovered.
Lower values indicate fewer findings to review per rediscovered CVE.

\section{Results}

\subsection{Different Models Find Different Bugs: Diversity Beats Repetition}

Figure~\ref{fig:curves} shows success@$k$ curves, and Table~\ref{tab:headline} lists four-pass results for all ten detector models with confidence intervals.
The best four-pass strict recall is 65/95 for GPT-5.6 luna in both context and no-context settings.
Five other model families reach at least 50 CVEs in one condition.
Moving from one pass to four adds approximately 12--20 CVEs per model.
Across both context conditions, 62\% of successful model--CVE--condition cells are detected in only one, two, or three of four identical passes.
On average, a single pass recovers 68\% of the CVEs demonstrated over four passes.

Model diversity increases rediscovery beyond repetition.
Using four scanner passes in total, one from each of the four leading models, covers an expected 68 distinct CVEs.
Four repeated passes of the strongest individual model, GPT-5.6 luna, cover 65.
The union of all forty context runs reaches 84/95 CVEs (85 counting both conditions), and twelve passes split across three inexpensive models find 70 CVEs for \$66, compared with 65 for \$128 from four GPT-5.6 luna passes.

\begin{table}[!b]
  \centering
  \caption{Strict rediscovery results for all ten detector models. Individual-model rows report pass@4 counts and descriptive 95\% Wilson intervals over the 95 benchmark CVEs. The portfolio row is an exploratory expectation over one pass per model.}
  \label{tab:headline}
  \begin{tabular}{lrrr}
    \toprule
    Model & Ctx & No ctx & Best result \\
    \midrule
    GPT-5.6 luna & 65 & 65 & 68.4\% (58.5--76.9) \\
    DeepSeek V4 Flash & 62 & 56 & 65.3\% (55.3--74.1) \\
    GPT-5.4 mini & 58 & 57 & 61.1\% (51.0--70.2) \\
    Gemini 3.5 Flash & 56 & 53 & 58.9\% (48.9--68.3) \\
    GPT-OSS 120B & 52 & 48 & 54.7\% (44.7--64.4) \\
    Qwen 3.6 35B A3B & 50 & 49 & 52.6\% (42.7--62.4) \\
    GPT-5.4 nano & 46 & 47 & 49.5\% (39.6--59.4) \\
    Gemini 2.5 Flash & 47 & 42 & 49.5\% (39.6--59.4) \\
    GPT-OSS 20B & 40 & 40 & 42.1\% (32.7--52.2) \\
    Mistral Small 4 & 38 & 36 & 40.0\% (30.7--50.1) \\
    \midrule
    Top-four portfolio$^\dagger$ & 68.0 & 68.2 & 71.8\% expected \\
    \bottomrule
  \end{tabular}
  \vspace{0.35em}

  \begin{minipage}{0.9\linewidth}
    \footnotesize
    $^\dagger$ One pass each from GPT-5.6 luna, DeepSeek V4 Flash, GPT-5.4 mini, and Gemini 3.5 Flash.
    Values average over all 256 combinations obtained by selecting one of four observed passes from each model; the portfolio analysis is exploratory.
  \end{minipage}
\end{table}

\subsection{Generated Context Is Not a Universal Gain}

At pass@4, context yields strict rediscovery in 514 of 950 model--CVE cells (54.1\%), compared with 493 of 950 (51.9\%) without context.
The paired difference is only +21 of 950 outcomes (2.2 percentage points; CVE-bootstrap 95\% interval $[-6,47]$), although context reduces the number of CVEs missed by every model from 16 to 11.
It also raises the number of findings (after deduplication) from 1,664 to 2,088.
A model with context therefore reports an average of 4.1 findings per rediscovered CVE, compared with 3.4 without context.
Most of the additional findings increase review work without improving strict recall.

Triage depth has a similarly small, model-dependent effect: using three triage reviews plus an arbiter instead of a single review changes the total number of rediscovered model--CVE outcomes only from 974 to 1,007; full results and caveats are in the supplement.

\subsection{Language Shapes Difficulty More Than Model Choice}

\begin{figure}[!htbp]
  \centering
  \includegraphics[width=0.98\linewidth]{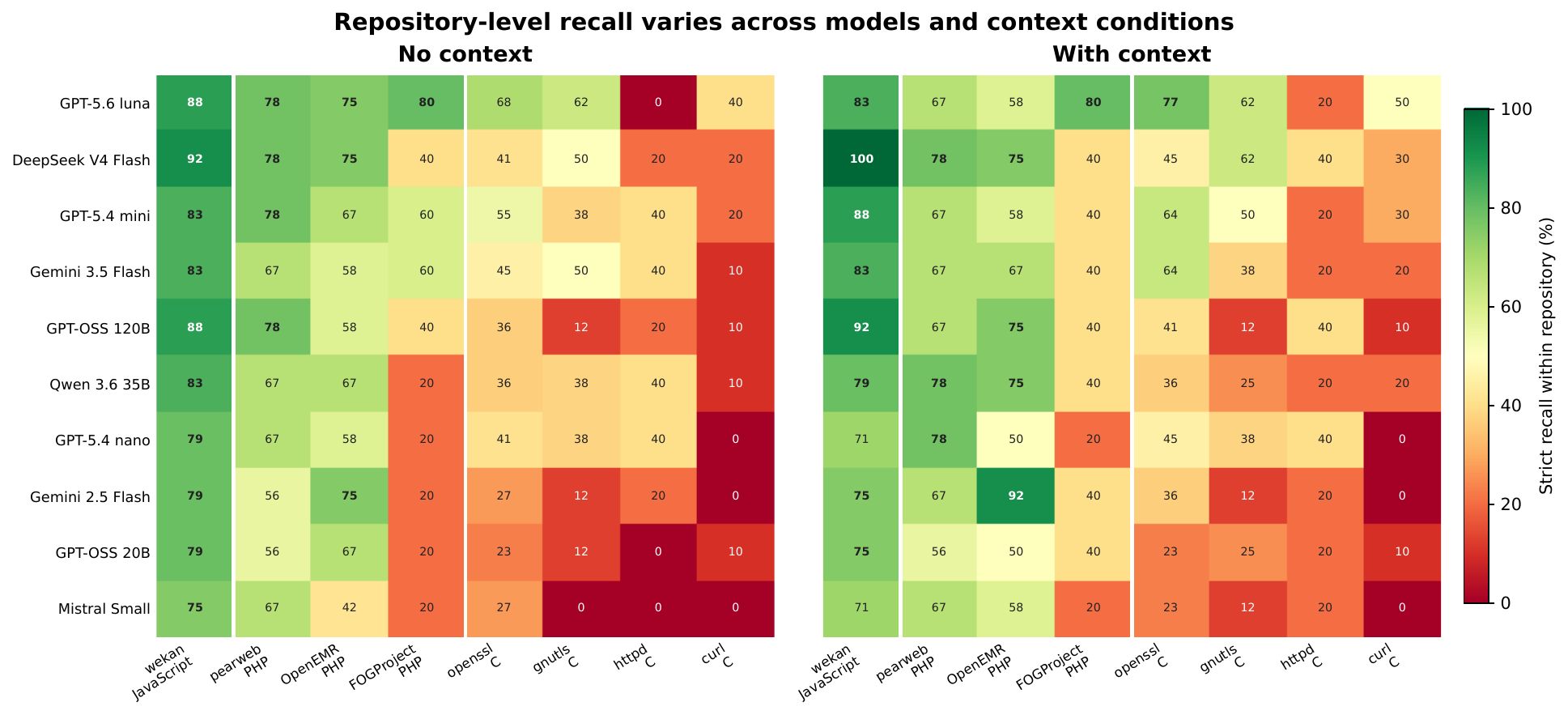}
  \caption{Strict pass@4 recall by model and repository. Rows are ordered by overall recall; repositories are grouped by language and ordered from easier to harder.}
  \label{fig:repo-heatmap}
\end{figure}

Figure~\ref{fig:repo-heatmap} aggregates strict rediscovery by repository, and Table~\ref{tab:language} collapses it by primary language.
With context, per-model recall is 71--100\% on JavaScript, 50--73\% on PHP, and only 16--62\% on C; nine of the eleven CVEs missed by every model are in C.
The pattern persists within vulnerability classes: among the 37 authorization and authentication CVEs (the only broad class spanning all three languages), an average of 8.1 models find each JavaScript CVE, 5.8 each PHP CVE, and 3.0 each C CVE.
Language explains more than twice as much deviance as vulnerability class in a binomial regression, and source size independently predicts difficulty ($\rho=-0.47$ overall; $-0.53$ within C).
Large, stateful C code appears harder regardless of the vulnerability mechanism.

\begin{table}[t]
  \centering
  \caption{Strict pass@4 rediscovery by primary language in the context condition. ``Median'' and ``best'' are over the ten detector models; ``missed by all'' counts CVEs found by no model (no-context counts in parentheses).}
  \label{tab:language}
  \begin{tabular}{lrrrrr}
    \toprule
    Language & CVEs & Median model & Best model & Found by any & Missed by all \\
    \midrule
    JavaScript & 24 & 19.5 (81\%) & 24 (100\%) & 24 & 0 \; (1) \\
    PHP & 26 & 16.5 (63\%) & 19 (73\%) & 24 & 2 \; (3) \\
    C & 45 & 14.0 (31\%) & 28 (62\%) & 36 & 9 \; (12) \\
    \bottomrule
  \end{tabular}
\end{table}

Median strict recall on the C CVEs is 31\% across the ten detector models.
With context, 21 CVEs are found by every model and 11 by none (22 and 16 without context); universally found cases include an OpenEMR SQL injection, while shared misses include subtle curl certificate-validation state bugs.

\section{Discussion}

The results show that vulnerability discovery is a systems problem, not a single model call \citep{fort2026mythos,aisle2026nano}.
The scaffold steers the model toward vulnerability reasoning, and triage keeps the volume of reported findings low enough for human review: even after four passes, the evaluated configurations report only 3--4 deduplicated findings per discovered CVE, a regime in which real signal is not drowned out.
This control is not free: repeated triage and arbitration dominate the attributed API cost.

Because generation is stochastic, a single scan understates what a scanner can do: across both context conditions, 62\% of successful model--CVE--condition cells here are found in only some of four identical passes.
Scanner evaluations should therefore report cumulative results over repeated runs, together with the volume of findings sent to review.

Performance also depends on the code under review.
Recall varies more by language and codebase than by vulnerability class: JavaScript and PHP application bugs are broadly rediscoverable, while large, stateful C infrastructure code is where the remaining headroom is concentrated.
We can only speculate why: in JavaScript and PHP the trust boundary is usually visible inside the target files, whereas the C cases turn on state assembled across many calls, so the evidence needed to justify a finding often lies outside what the detector sees.

The benchmark is also well suited for future fine-tuning experiments because raw traces include prompts, model responses, findings, and judge labels.
Fine-tuning may reduce variance and schema failures; it may also fit the model too closely to the prompt style or judge preferences.

These conclusions should be read within the benchmark's boundaries.
\hof\ is AISLE-conditioned and source-conditioned, and its eight repositories are not representative of all software.
It samples one vendor's slice of the broader agentic-CVE population that \citet{bvi2026agentic} catalog, and readers who want to know how representative that slice is should compare against their census rather than take our selection on trust.
Every task is positive and supplies ground-truth-derived target files, so the study estimates target-conditioned recall, not whole-repository localization, precision, or false-positive rate; unmatched findings are unverified, not false positives.
All ten detector backbones share one scaffold, prompt family, triage design, and judge.
The results therefore cover ten configurations of a single analyzer and should not be interpreted as a comparison of ten independent products.
The CVEs are public, so training-data contamination cannot be excluded, and the judge and semantic deduplication are automated and have not been human-audited.
The judge also shares a vendor family with several of the detectors, so a same-family scoring preference cannot be ruled out; the judge is blinded to detector identity, but blinding does not remove stylistic affinity between a model and reports written in its own family's idiom.

\section{Ethics and Competing Interests}

\hof\ comprises public CVEs discovered and disclosed by AISLE, a developer of commercial AI-based vulnerability-detection systems.
All authors were employed by or affiliated with AISLE during the work reported here, and both the benchmark population and the reference scanner design originate there.
Readers should treat \hof\ as a vendor-constructed benchmark: we have therefore released the dataset, the scanner-visible manifest, and the raw run artifacts so that the selection and scoring decisions can be audited independently.
This paper contains no exploit code or unpublished vulnerability details.

\section*{Acknowledgments}

We thank the maintainers of OpenSSL, curl, GnuTLS, Apache httpd, pearweb, FOGProject, OpenEMR, and WeKan for their work triaging and fixing the vulnerabilities that make up this benchmark.

\bibliographystyle{plainnat}
\bibliography{references}

@misc{aisle2026nano,
  author = {Fort, Stanislav},
  title = {System Over Model: Zero-Day Discovery at the Jagged Frontier},
  year = {2026},
  month = apr,
  howpublished = {AISLE blog},
  url = {https://aisle.com/blog/system-over-model-zero-day-discovery-at-the-jagged-frontier},
  note = {Published April 14, 2026}
}

@misc{fort2026mythos,
  author = {Fort, Stanislav},
  title = {{AI} Cybersecurity After {Mythos}: The Jagged Frontier},
  year = {2026},
  month = apr,
  howpublished = {AISLE blog},
  url = {https://aisle.com/blog/ai-cybersecurity-after-mythos-the-jagged-frontier},
  note = {Published April 7, 2026}
}

@misc{aisle2026cves,
  author = {{AISLE}},
  title = {{CVE} Discoveries},
  year = {2026},
  howpublished = {Public Hall of Fame},
  url = {https://aisle.com/cve-discoveries},
  note = {Accessed July 2026}
}

@misc{bvi2026agentic,
  author = {Villa, Corban and Kaviani, Darya and Cheriton, Alice and Popa, Raluca Ada},
  title = {The Agentic Vulnerability Coverage Map},
  year = {2026},
  howpublished = {Berkeley Vulnerability Initiative, UC Berkeley EECS},
  url = {https://vuln.cs.berkeley.edu/},
  note = {Accessed July 2026}
}

@misc{sun2026drea,
  author = {Sun, Mingyang and Meng, Guozhu},
  title = {{DREA}: Decoupled Reasoning and Exploration Agents for Repository-Level Vulnerability Detection},
  year = {2026},
  howpublished = {arXiv preprint arXiv:2607.13439},
  url = {https://arxiv.org/abs/2607.13439}
}

@misc{realvuln2026,
  author = {Pellew, John and Raza, Faizan},
  title = {{RealVuln}: Benchmarking Rule-Based, General-Purpose {LLM}, and Security-Specialized Scanners on Real-World Code},
  year = {2026},
  howpublished = {arXiv preprint arXiv:2604.13764},
  url = {https://arxiv.org/abs/2604.13764}
}

@inproceedings{bountybench2025,
  author = {Zhang, Andy K. and others},
  title = {{BountyBench}: Dollar Impact of {AI} Agent Attackers and Defenders on Real-World Cybersecurity Systems},
  booktitle = {Advances in Neural Information Processing Systems},
  volume = {38},
  year = {2025}
}

@inproceedings{secbench2025,
  author = {Lee, Hwiwon and Zhang, Ziqi and Lu, Hanxiao and Zhang, Lingming},
  title = {{SEC-bench}: Automated Benchmarking of {LLM} Agents on Real-World Software Security Tasks},
  booktitle = {Advances in Neural Information Processing Systems},
  volume = {38},
  year = {2025}
}

@inproceedings{jitvul2025,
  author = {Yildiz, Alperen and Teo, Sin G. and Lou, Yiling and Feng, Yebo and Wang, Chong and Divakaran, Dinil Mon},
  title = {Benchmarking {LLMs} and {LLM}-based Agents in Practical Vulnerability Detection for Code Repositories},
  booktitle = {Proceedings of the 63rd Annual Meeting of the Association for Computational Linguistics (Volume 1: Long Papers)},
  pages = {30848--30865},
  publisher = {Association for Computational Linguistics},
  address = {Vienna, Austria},
  doi = {10.18653/v1/2025.acl-long.1490},
  year = {2025}
}

@inproceedings{primevul2025,
  author = {Ding, Yangruibo and Fu, Yanjun and Ibrahim, Omniyyah and Sitawarin, Chawin and Chen, Xinyun and Alomair, Basel and Wagner, David A. and Ray, Baishakhi and Chen, Yizheng},
  title = {Vulnerability Detection with Code Language Models: How Far Are We?},
  booktitle = {Proceedings of the 47th IEEE/ACM International Conference on Software Engineering},
  pages = {1729--1741},
  publisher = {IEEE},
  doi = {10.1109/ICSE55347.2025.00038},
  year = {2025}
}

@inproceedings{devign2019,
  author = {Zhou, Yaqin and Liu, Shangqing and Siow, Jingkai and Du, Xiaoning and Liu, Yang},
  title = {Devign: Effective Vulnerability Identification by Learning Comprehensive Program Semantics via Graph Neural Networks},
  booktitle = {Advances in Neural Information Processing Systems},
  volume = {32},
  pages = {10197--10207},
  year = {2019}
}

@inproceedings{bigvul2020,
  author = {Fan, Jiahao and Li, Yi and Wang, Shaohua and Nguyen, Tien N.},
  title = {A {C/C++} Code Vulnerability Dataset with Code Changes and {CVE} Summaries},
  booktitle = {Proceedings of the 17th International Conference on Mining Software Repositories},
  pages = {508--512},
  publisher = {ACM},
  doi = {10.1145/3379597.3387501},
  year = {2020}
}

@article{juliet,
  author = {Boland, Jr., Frederick E. and Black, Paul E.},
  title = {The Juliet 1.1 {C/C++} and Java Test Suite},
  journal = {Computer},
  volume = {45},
  number = {10},
  pages = {88--90},
  doi = {10.1109/MC.2012.345},
  year = {2012}
}

@techreport{sard,
  author = {Black, Paul E.},
  title = {The Software Assurance Reference Dataset ({SARD})},
  institution = {National Institute of Standards and Technology},
  number = {NIST IR 8561},
  doi = {10.6028/NIST.IR.8561},
  year = {2025}
}

@misc{snykvulnbench2026,
  author = {Tal, Liran and Kloos, Johannes and Rudich, Arsenii and Thoemmes, Stephen and Nair, Manoj},
  title = {Snyk VulnBench {JS} 1.0: Can {LLMs} Find the Same Bugs Twice?},
  year = {2026},
  month = jun,
  howpublished = {arXiv preprint arXiv:2606.15762},
  url = {https://arxiv.org/abs/2606.15762}
}

@inproceedings{simecek2026silent,
  author = {Simecek, Petr and Cadek, Vaclav and Buran, Michal},
  title = {Silent Failures in {LLM} Vulnerability Detection: The Expensive Model Trap},
  booktitle = {Failure Modes in Agentic AI (FAGEN), ICML 2026 Workshop},
  year = {2026},
  url = {https://openreview.net/forum?id=0xklkDx3U1}
}

@inproceedings{buran2026benchmarks,
  author = {Buran, Michal and Cadek, Vaclav},
  title = {Benchmarks as Random Variables---Modeling Overdispersion in {LLM} Evaluation},
  booktitle = {Structured Probabilistic Inference \& Generative Modeling (SPIGM), ICML 2026 Workshop},
  year = {2026},
  url = {https://openreview.net/forum?id=g5jezzrnct}
}

@book{rasch1960,
  author = {Rasch, Georg},
  title = {Probabilistic Models for Some Intelligence and Attainment Tests},
  publisher = {Danish Institute for Educational Research},
  address = {Copenhagen, Denmark},
  year = {1960}
}

@book{lord1980,
  author = {Lord, Frederic M.},
  title = {Applications of Item Response Theory to Practical Testing Problems},
  publisher = {Lawrence Erlbaum Associates},
  address = {Hillsdale, NJ},
  isbn = {0-89859-006-X},
  year = {1980}
}

@book{reckase2009,
  author = {Reckase, Mark D.},
  title = {Multidimensional Item Response Theory},
  publisher = {Springer},
  address = {New York, NY},
  doi = {10.1007/978-0-387-89976-3},
  year = {2009}
}

@article{vehtari2017,
  author = {Vehtari, Aki and Gelman, Andrew and Gabry, Jonah},
  title = {Practical Bayesian Model Evaluation Using Leave-One-Out Cross-Validation and WAIC},
  journal = {Statistics and Computing},
  volume = {27},
  number = {5},
  pages = {1413--1432},
  doi = {10.1007/s11222-016-9696-4},
  year = {2017}
}

@article{hoffman2014,
  author = {Hoffman, Matthew D. and Gelman, Andrew},
  title = {The No-U-Turn Sampler: Adaptively Setting Path Lengths in Hamiltonian Monte Carlo},
  journal = {Journal of Machine Learning Research},
  volume = {15},
  number = {47},
  pages = {1593--1623},
  year = {2014}
}

\clearpage
\appendix
\section{Supplement: Benchmark Format}

Each scanner task includes repository URL, scan commit, language, and target files.
The evaluator-only ground truth contains CVE ID, vulnerable paths, line ranges, description, code snippet, severity, category, CWE, and fix metadata.
Table~\ref{tab:schema} summarizes the field separation; the scanner-visible manifest is leakage-tested against the ground truth.
The benchmark and reference harness are available at:
\begin{itemize}[leftmargin=1.2em]
  \item Hugging Face dataset: \url{https://huggingface.co/datasets/aisleinc/HoF-Bench}
  \item GitHub repository: \url{https://github.com/weareaisle/HoF-Bench}
\end{itemize}

\begin{table}[h]
\centering
\caption{Benchmark field separation. Ground-truth fields are never included in scanner prompts.}
\label{tab:schema}
\begin{tabular}{lp{0.62\linewidth}}
\toprule
View & Fields \\
\midrule
Scanner-visible & task ID, repository URL and ID, pinned commit, language, target files, scope mode \\
Evaluator-only & CVE ID, expected and related paths, line range, title, mechanism, impact, code snippet, CWE/category/severity, fix metadata \\
Scanner finding & title, normalized path, line range, root cause, attacker control, impact, evidence, confidence \\
Judge verdict & strict match, partial, no match, or insufficient evidence; four identity dimensions and rationale \\
\bottomrule
\end{tabular}
\end{table}

\section{Supplement: Effect of Triage Depth}

Each finding is reviewed by up to three sequential triage rounds; each round returns a verdict (valid, invalid, or uncertain) and sees the prior rounds' reviews.
When multiple rounds are used, an arbiter model reviews the finding, source, and all round outputs and issues the final decision based on code evidence rather than majority vote.
All per-round verdicts are stored in the raw call records, which allows a retrospective replay: under a one-round rule a finding survives if round one voted valid, and under a two-round rule if valid votes outnumber invalid votes in the first two rounds.
Simulated survivors that the full protocol rejected had never reached the judge, so we scored all 645 such findings with the same blinded strict judge (about \$4 of additional judge calls); 70 were strict matches and 28 partial.
Two caveats apply.
Later rounds were generated with knowledge of earlier reviews, so a true one-round run might phrase its reviews differently, and we do not simulate an arbiter for the two-round rule.
The replay therefore estimates the filtering behavior of the recorded reviews, not a full counterfactual rerun.
Per-model results are in \code{results/triage\_rounds\_summary.csv} in the artifact package.

\section{Supplement: One-Dimensional Reliability Model}
\label{sec:reliability-model}

The analysis follows two recent AISLE studies that treat a benchmark score as a random variable rather than a fixed property of a model.
\emph{Benchmarks as Random Variables} \citep{buran2026benchmarks} makes this point in a hierarchical Rasch setting \citep{rasch1960,lord1980} that separates model ability from item difficulty and adds a single overdispersion parameter, and finds that conventional reporting understates uncertainty even when it ranks models correctly.
\emph{Silent Failures in LLM Vulnerability Detection} \citep{simecek2026silent} extends the framing to multidimensional item response theory \citep{reckase2009}, resolving detection failure into several latent axes with a three-factor Beta-Binomial fit over 50 CVEs and 50 models.
Its most relevant conclusion here is operational: reaching a fixed reliability target by repeating the single strongest model cost roughly \$41 on that benchmark, while a mixed portfolio of cheap models reached it for under \$1.
That is the same diversity effect we measure on a different CVE population.

For \hof\ we fit a deliberately simpler model that asks only one question: how strong is a scanner configuration relative to the difficulty of a CVE?
With four passes per cell there is too little repetition to recover several latent security capabilities, so we do not try.

\begin{figure}[h]
  \centering
  \includegraphics[width=0.85\linewidth]{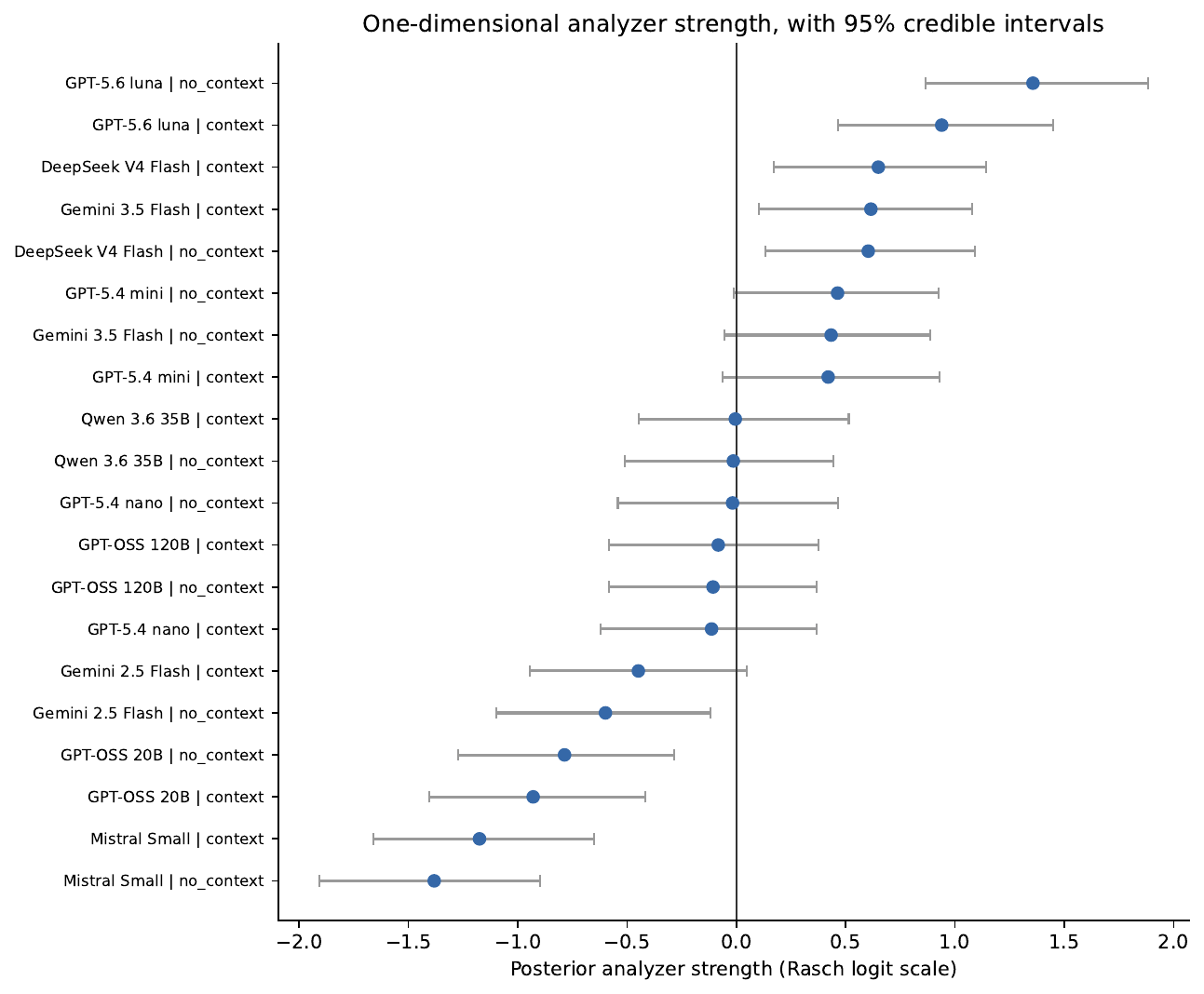}
  \caption{Posterior analyzer strength under the one-dimensional Rasch Beta-Binomial model. Intervals are 95\% credible intervals; zero is the population-centered ability scale, not a performance threshold.}
  \label{fig:irt-strength}
\end{figure}

Let $y_{ij}$ be the number of strict detections by scanner configuration $i$ on CVE $j$ across $n_{ij}$ passes.
We model
\begin{align}
  \operatorname{logit}(p_{ij}) &= \alpha + \theta_i - \delta_j,\\
  y_{ij} &\sim \operatorname{BetaBinomial}\bigl(n_{ij},\, \kappa p_{ij},\, \kappa(1-p_{ij})\bigr),
\end{align}
where $\theta_i$ is scanner strength, $\delta_j$ is CVE difficulty, and both are centered at zero, so the global intercept $\alpha$ is the log-odds that an average configuration detects an average CVE.
The concentration $\kappa$ captures overdispersion across repeated passes: smaller $\kappa$ means passes on the same cell agree less than independent coin flips would predict.
We fit this Rasch mean structure and a 2PL variant, which adds a per-CVE discrimination parameter multiplying $\theta_i$, under both binomial and Beta-Binomial likelihoods, using four NUTS chains with 1,000 warmup iterations and 1,000 posterior draws per chain \citep{hoffman2014}.
Table~\ref{tab:irt-comparison} reports the PSIS-LOO comparison \citep{vehtari2017}: both Beta-Binomial variants beat their binomial counterparts by over 100 ELPD, confirming that repeated passes are more correlated than binomial sampling allows, while the extra 2PL discrimination parameter adds nothing ($\Delta$ELPD $=5.6$, dSE $=6.3$).
Figure~\ref{fig:irt-strength} shows posterior analyzer strengths under the preferred Rasch Beta-Binomial model.

\begin{table}[h]
\centering
\caption{One-dimensional reliability-model comparison. ELPD differences are relative to the preferred Rasch Beta-Binomial model.}
\label{tab:irt-comparison}
\begin{tabular}{lrrr}
\toprule
Model & ELPD & $\Delta$ELPD & dSE \\
\midrule
Rasch Beta-Binomial & $-1769.1$ & 0.0 & --- \\
2PL Beta-Binomial & $-1774.7$ & 5.6 & 6.3 \\
2PL Binomial & $-1879.5$ & 110.4 & 21.1 \\
Rasch Binomial & $-1924.1$ & 155.0 & 21.4 \\
\bottomrule
\end{tabular}
\end{table}

\clearpage
\section{Supplementary Figures}

\vspace*{\fill}
\vspace*{\fill}

\begin{figure}[h]
  \centering
  \includegraphics[width=0.85\linewidth]{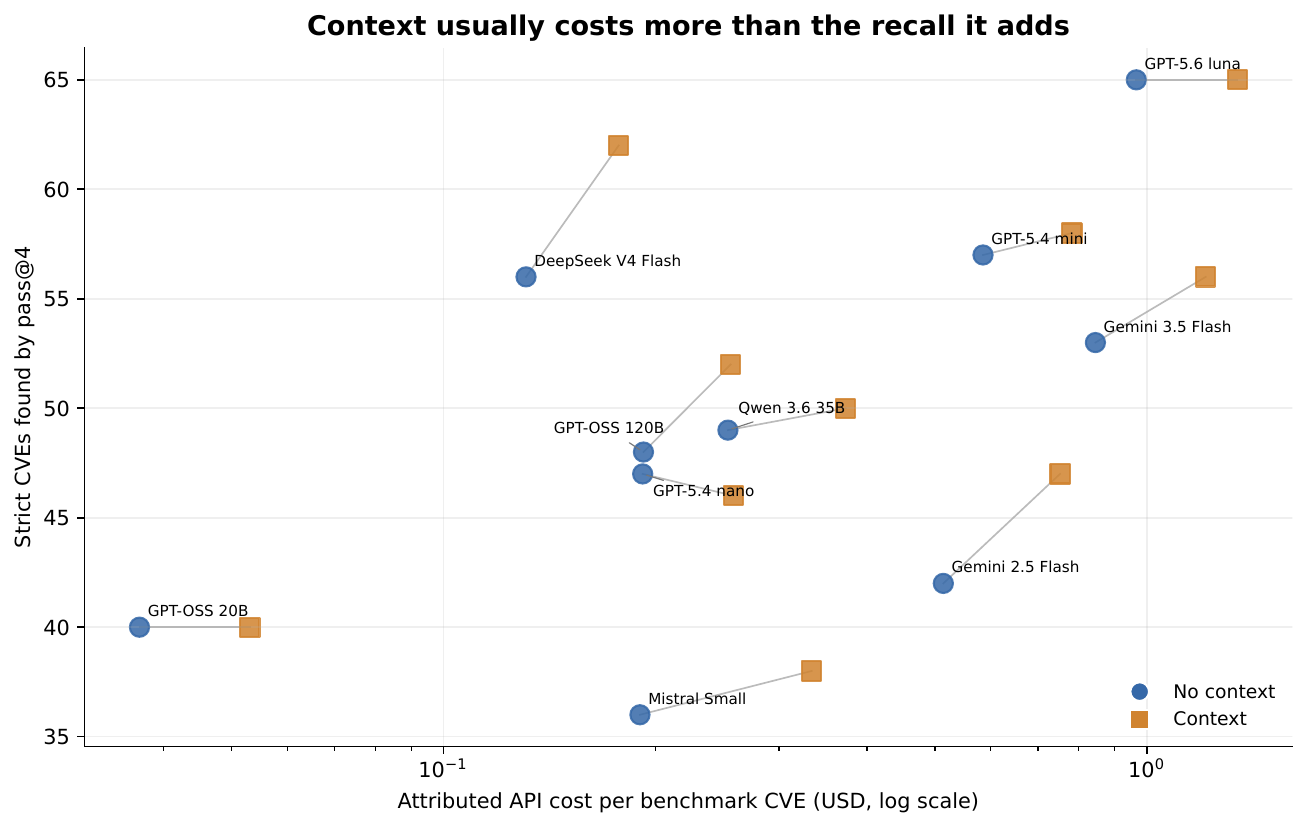}
  \caption{Attributed API cost per benchmark CVE versus strict rediscovery. This is an experimental accounting view, not a production pricing claim.}
  \label{fig:supp-recall-cost}
\end{figure}

\end{document}